\documentclass[10pt,twocolumn,letterpaper]{article}

\usepackage[dvipsnames, table, svgnames, rgb]{xcolor}
\usepackage{colortbl}

\usepackage{booktabs} 
\usepackage{pifont}
\newcommand{\cmark}{\ding{51}} 

\usepackage{multirow}
\usepackage{arydshln}
\usepackage{comment}

\usepackage{cuted}
\usepackage{capt-of}

\usepackage{listings}

\definecolor{codegray}{rgb}{0.5,0.5,0.5}
\definecolor{codepurple}{rgb}{0.58,0,0.82}
\definecolor{backcolor}{rgb}{0.95,0.95,0.92}

\lstdefinestyle{mystyle}{
    backgroundcolor=\color{backcolor},   
    commentstyle=\color{codegray},       
    keywordstyle=\color{blue},           
    numberstyle=\tiny\color{codegray},   
    stringstyle=\color{codepurple},      
    basicstyle=\ttfamily\footnotesize,   
    breakatwhitespace=false,             
    breaklines=true,                     
    captionpos=b,                        
    keepspaces=true,                     
    numbers=left,                        
    numbersep=5pt,                       
    showspaces=false,                    
    showstringspaces=false,              
    showtabs=false,                      
    tabsize=2                            
}

\usepackage[accsupp]{axessibility} 

\usepackage[pagenumbers]{cvpr} 

\definecolor{cvprblue}{rgb}{0.21,0.49,0.74}
\usepackage[pagebackref,breaklinks,colorlinks,citecolor=cvprblue]{hyperref}

\def\paperID{14418} 
\def\confName{CVPR}
\def\confYear{2025}

\title{CoMapGS: Covisibility Map-based Gaussian Splatting\\for Sparse Novel View Synthesis}

\author{Youngkyoon Jang and Eduardo P\'erez-Pellitero\\
Huawei Noah's Ark Lab, London, UK
}

\begin{document}
\maketitle

\begin{strip}\centering
\vspace*{-4em}
\includegraphics[width=\textwidth]{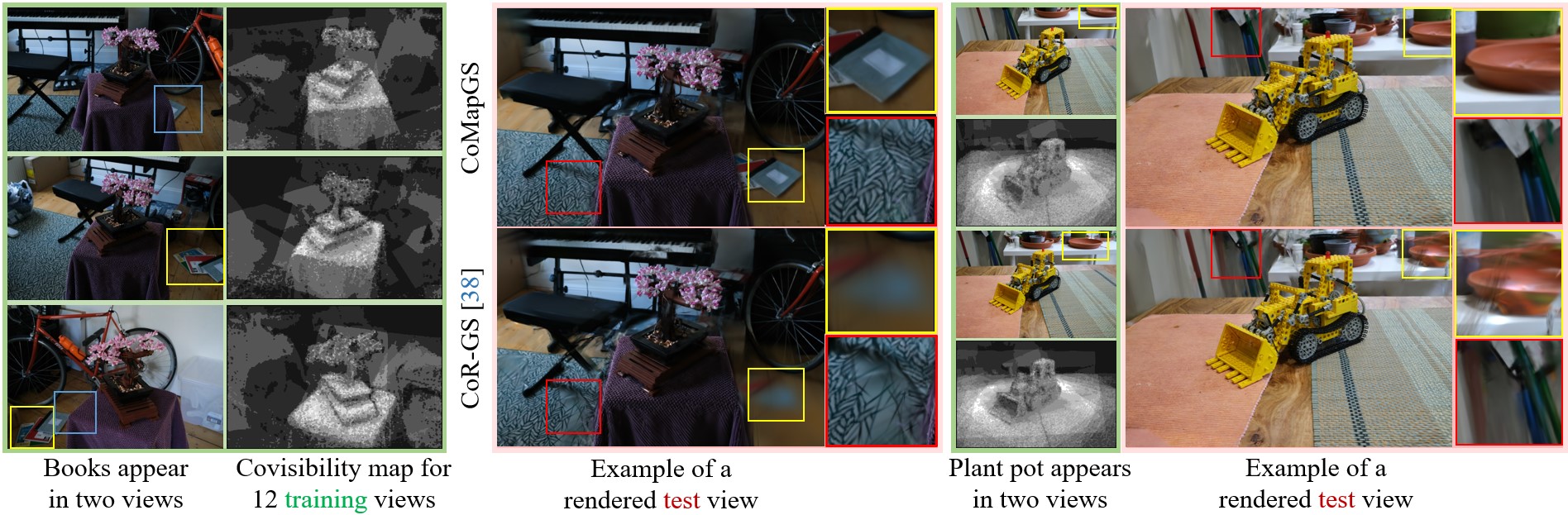}
\captionof{figure}{\textbf{Main idea.} Conventional sparse view synthesis methods using 3DGS~\citep{Zhang_2024_ECCV_CorGS} prioritize frequently captured regions, shown as the brightest areas in the covisibility map, resulting in better reconstruction in these regions but missing details in sparsely captured areas. By using a covisibility map, we guide sparse 3DGS methods to focus on underrepresented regions, enhancing sparse novel view synthesis.
\label{fig:teaser}}
\end{strip}

\begin{abstract}
We propose Covisibility Map-based Gaussian Splatting (CoMapGS), designed to recover underrepresented sparse regions in sparse novel view synthesis. CoMapGS addresses both high- and low-uncertainty regions by constructing covisibility maps, enhancing initial point clouds, and applying uncertainty-aware weighted supervision using a proximity classifier. Our contributions are threefold: (1) CoMapGS reframes novel view synthesis by leveraging covisibility maps as a core component to address region-specific uncertainty; (2) Enhanced initial point clouds for both low- and high-uncertainty regions compensate for sparse COLMAP-derived point clouds, improving reconstruction quality and benefiting few-shot 3DGS methods; (3) Adaptive supervision with covisibility-score-based weighting and proximity classification achieves consistent performance gains across scenes with varying sparsity scores derived from covisibility maps. Experimental results demonstrate that CoMapGS outperforms state-of-the-art methods on datasets including Mip-NeRF 360 and LLFF.\footnote{Project page: \href{https://youngkyoonjang.github.io/projects/comapgs/}{youngkyoonjang.github.io/projects/comapgs}}
\end{abstract}    
\section{Introduction}
\label{sec:intro}

Recent advances in novel view synthesis have made significant progress in overcoming various challenges, such as generalized model building~\citep{Tanay_2024_CVPR, Zhu_2024_ECCV_CaesarNeRF}, view synthesis in unbounded scenes~\citep{barron2021mipnerf}, fast training~\citep{fan2024instantsplat, Fang_ECCV_2024_MiniSplatting}, and sparse view~\citep{Zhu_2024_ECCV_FSGS, Zhang_2024_ECCV_CorGS}, following the emergence of seminal methodologies such as Neural Radiance Fields (NeRF)~\citep{mildenhall2020nerf} and 3D Gaussian Splatting (3DGS)~\citep{kerbl3Dgaussians}. While dense-view approaches now achieve photo-realistic fidelity, sparse-view synthesis methods often suffer from unrealistic artifacts due to the fundamental shape-radiance ambiguity—\textit{i.e.}, models can overfit to image observations because of implausible geometry, a problem that is exacerbated when training views are limited. Despite the limitations of available training view images, recent studies have proposed various approaches—such as generalized novel view synthesis methods~\citep{Yu_2021_CVPR_PixelNeRF, Chen_2021_ICCV_MVSNeRF, kulhanek2022viewformer, Cong_2023_ICCV_GNT_MOVE, Tanay_2024_CVPR, Zhu_2024_ECCV_CaesarNeRF}, diffusion-based techniques~\citep{Wu_2024_CVPR, Zhou_2023_CVPR_SparseFusion}, updated densification strategies~\citep{Zhu_2024_ECCV_FSGS, Zhang_2024_ECCV_CorGS}, and the use of pseudo ground truth~\citep{Wu_2024_CVPR, Zhu_2024_ECCV_FSGS, Zhang_2024_ECCV_CorGS}—to improve performance in sparse view synthesis. While recent approaches, particularly those using 3DGS, have significantly improved performance in sparse view synthesis by leveraging rapid point-cloud-based rasterization, efficient gradient-descent supervision, and multiview geometry, several open challenges remain:

\begin{itemize}
\item {\textbf{Region-wise imbalanced supervision:}} Conventional novel view synthesis methods prioritize highly covisible regions due to their multiview presence, effectively weighting them more while under-supervising mono-view regions that do not have multiview constraints.
\item {\textbf{Sparse point cloud initialization:}} Conventional sparse point clouds, generated through feature-based keypoint matching (\eg, SIFT~\citep{LoweIJCV04SIFT}) in COLMAP~\citep{schoenberger2016sfm}, lack the geometric detail that dense view settings offer, especially when the number of training images is limited.
\item {\textbf{High-uncertainty regions:}} Methods that rely heavily on multiview geometry often struggle with mono-view regions where pixels do not appear in other views, posing a critical challenge in sparse view synthesis.
\end{itemize}

\begin{figure*}
\centering
\includegraphics[width=0.999\linewidth]{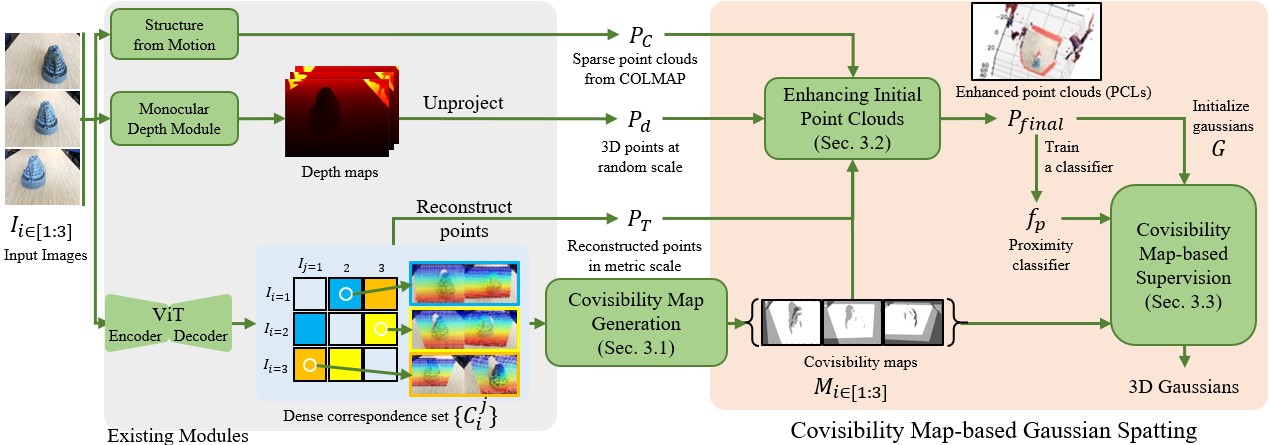}
\caption{Overview of the proposed CoMapGS. CoMapGS leverages existing modules, including Structure-from-Motion (SfM), monocular depth, and dense correspondence prediction, to produce preliminary outputs. Covisibility maps are generated and used alongside module outputs to enhance initial point clouds, which initialize Gaussians for CoMapGS. During training, covisibility map-based weighted supervision with the proximity classifier applies varied strengths to Gaussians based on region-specific covisibility.
}
\label{fig:architecture_overview}
\end{figure*}

In response to these challenges, we propose CoMapGS, a Covisibility Map-based Gaussian Splatting method that reframes sparse view synthesis by utilizing per-image, pixel-wise covisibility counts (\ie, the covisibility map) for adaptive supervision and initial point cloud enhancement, effectively addressing both low- and high-uncertainty regions, as shown in Fig.~\ref{fig:architecture_overview}. CoMapGS updates initial point clouds and applies adaptive supervision, using a proximity classifier to regularize plausible geometry in high-uncertainty (\ie, mono-view) regions for enhanced supervision. To effectively manage regions with varying covisibility levels, we construct a covisibility map by accumulating correspondence counts for each pixel from the output of the state-of-the-art dense correspondence prediction model, MASt3R~\citep{mast3r_arxiv24}. Additionally, we recover initial point clouds across all regions, including mono-view areas, by learning the correlation between the scale of monocular depth estimation and the point clouds registered using already-optimized COLMAP camera poses, enabling precise rescaling and alignment within the same coordinate system. The updated point clouds are then integrated into CoMapGS to initialize the first set of Gaussians, which are rendered and refined throughout the training process.

Our proposed CoMapGS can be applied to most sparse view synthesis methods built on the conventional 3DGS framework. Starting with our updated initial point clouds, we supervise CoMapGS by incorporating an additional proximity loss term into the baseline objective function (\eg, FSGS~\citep{Zhu_2024_ECCV_FSGS}, CoR-GS~\citep{Zhang_2024_ECCV_CorGS}). This proximity loss is derived from our proposed proximity MLP classifier, which predicts the closeness of each point to the original geometry. Leveraging the covisibility map, we apply adaptive weights to the Gaussians that compose the final proximity term, assigning greater weight to Gaussians projected onto mono-view regions and reduced weight to those in multiview regions. Additionally, we adjust supervision strength for Gaussians outside the view frustum based on scene-specific covisibility scores, ensuring they contribute effectively to training even if they are not directly represented in rendered images.

The main contributions of the CoMapGS are threefold:

\begin{enumerate}
\item {\textbf{Covisibility Map-based learning:}} CoMapGS reframes sparse view synthesis by leveraging covisibility maps for adaptive supervision, enabling balanced guidance across different uncertainty levels within the scene.
\item {\textbf{Enhanced point cloud initialization:}} By applying learned scaling parameters between monocular depth estimates and triangulated points registered using the given COLMAP camera poses, CoMapGS densely enhances initial point cloud quality, increasing the potential to densify sparse and underrepresented regions, even in mono-view areas.
\item {\textbf{Weighted supervision using covisibility map:}} This strategy adaptively reinforces high- and low-uncertainty regions, strengthening supervision based on region-specific covisibility levels and stabilizing training with learned geometric constraints that extend beyond the view frustum.
\end{enumerate}

\noindent To the best of our knowledge, CoMapGS is the first approach to explicitly address and recover high-uncertainty regions in sparse view synthesis, effectively balancing supervision across regions with varying uncertainty levels.

\section{Related Works}
\label{sec:related_works}

\textbf{Sparse novel view synthesis.} The growing interest in reconstructing spatial photographs from minimal input image sets has driven the development of various methods aimed at generating novel views from just a few shots, utilizing approaches based on NeRF~\citep{Niemeyer2021Regnerf, Jain_2021_ICCV_DietNeRF, Roessle_2022_CVPR_DepthPrior, Deng_2022_CVPR, Kim_2022_CVPR, Yang2023FreeNeRF, Song_NeurIPS_2023_DaRF, Wynn_2023_CVPR_DiffusionNeRF, Wang_2023_ICCV_SparseNeRF, Wu_2024_CVPR} and 3DGS~\citep{xiong2023sparsegs, Li_2024_CVPR_DNGaussian, paliwal2024coherentgs, Zhu_2024_ECCV_FSGS, Zhang_2024_ECCV_CorGS, Xu_ECCV_2024_MVPGS}. Due to the inherent challenges of sparse view synthesis, such as the limited number of training views, NeRF-based studies have primarily focused on proposing regularization techniques with depth ranking~\citep{Deng_2022_CVPR, Roessle_2022_CVPR_DepthPrior, Kwak_ICML_2023_GeCoNeRF, Song_NeurIPS_2023_DaRF, Wynn_2023_CVPR_DiffusionNeRF, Wang_2023_ICCV_SparseNeRF}, frequency-based loss~\citep{Yang2023FreeNeRF}, semantic consistency with visual encoders~\citep{Jain_2021_ICCV_DietNeRF}, between-view smoothness~\citep{Kim_2022_CVPR}, or uncertainty-aware adaptive loss~\citep{Xu_ECCV_2024_ARNeRF} for both seen and unseen views~\citep{Niemeyer2021Regnerf, Jain_2021_ICCV_DietNeRF, Kwak_ICML_2023_GeCoNeRF, Wu_2024_CVPR}. Some methods have also addressed uncertainty issues~\citep{Kim_2022_CVPR, Yang2023FreeNeRF}, particularly for high-uncertainty mono-view or unseen regions. However, these methods either disregard~\citep{Kim_2022_CVPR} or penalize~\citep{Yang2023FreeNeRF} uncertain areas, rather than fully utilizing the information contained within. 

While recent 3DGS-based methods continue to rely on regularization techniques, such as depth constraints~\citep{xiong2023sparsegs, Li_2024_CVPR_DNGaussian, paliwal2024coherentgs, Zhu_2024_ECCV_FSGS, Xu_ECCV_2024_MVPGS}, some have additionally introduced novel densification strategies~\citep{Zhu_2024_ECCV_FSGS, Zhang_2024_ECCV_CorGS} and explored \textbf{enhancing initial point clouds}~\citep{paliwal2024coherentgs, Xu_ECCV_2024_MVPGS, fan2024instantsplat, chen2024dustgs}, capitalizing on the technical characteristics of 3DGS. These methods primarily focus on improving effective multiview regions~\citep{gao2022dynamic}, often biased towards well-represented regions in the training views, rather than addressing recovery strategies for high-uncertainty mono-view regions. To overcome the inherent limitations of sparse training views, several prior works have explored generating and utilizing pseudo-ground truths for novel viewpoints via mono-depth estimation~\citep{Zhu_2024_ECCV_FSGS}, forward warping with reversed bilinear sampling~\citep{Xu_ECCV_2024_MVPGS}, or diffusion models with CLIP embedding~\citep{Wu_2024_CVPR}. While these approaches compensate for the lack of training views, they still face unique challenges in recovering unseen or mono-view regions from the original training view images, potentially misleading the learning process in high-uncertainty areas. 

\textbf{Covisibility-aware supervision.} Beginning with DyCheck~\citep{gao2022dynamic}, which emphasizes the importance of covisibility in multiview regions to address fair evaluation in dynamic novel view synthesis, recent studies have explored integrating covisibility information into supervision strategies. NeRFVS~\citep{YangCVPR2023NeRFVS} uses a view coverage map to enforce stricter regularization in low-coverage (covisible) regions, but it is primarily suited for dense view synthesis with video input, where extensive multiview overlap is expected. Similarly, InfoNorm~\citep{Wang_ECCV_2024_InfoNorm} leverages mutual information among surface normals at highly correlated scene points to regularize geometric modeling. However, InfoNorm captures covisible areas within local subregions in structured scenes (\eg, rooms) rather than using a comprehensive covisibility map, which limits its applicability for addressing both multiview and monoview regions. While these methods demonstrate the value of covisibility in supervision, they lack a framework that addresses the challenges of sparse novel view synthesis for natural scenes, including outdoor environments, through a pixel-wise covisibility map.

\section{Covisibility Map-based Gaussian Splatting}
\label{sec:method}

The proposed CoMapGS framework is illustrated in Fig.~\ref{fig:architecture_overview}. CoMapGS builds on the latest 3DGS-based sparse novel view synthesis method, CoR-GS~\citep{Zhang_2024_ECCV_CorGS}, with two key improvements: (1) updating initial point clouds specifically for sparse and mono-view regions and (2) applying adaptive supervision across varying covisibility regions using a covisibility map and proximity MLP classifier. We begin by generating covisibility maps using the dense correspondence prediction model MASt3R~\citep{mast3r_arxiv24}, refining it with morphological operations (\eg, erosion and dilation) to support these strategies effectively. 

\subsection{Covisibility Map Generation}
\label{subsec:CoMap_genration}

We generate a covisibility map by identifying dense correspondences between each training view image and all other training view images. Given a set of training view images $T = \{I_1, I_2, \dots, I_n\}$, where $n$ is the total number of training views, we compute a covisibility map $M_i$ for each view $I_i$. For every pair ($I_i, I_{\{j\}}$) with $j \neq i$, dense correspondences $C_i^j$ are predicted by a dense correspondence prediction function $f_{\text{cor}}$, where $C_i^j = f_{\text{cor}}(I_i, I_j)$. The covisibility map $M_i$ is initialized as a zero matrix of size $\mathsf{W} \times \mathsf{H}$ and defined as:
\begin{equation}
    M_i(x, y) = \sum_{j=1, j \neq i}^{n} \delta_{(x, y) \in \mathcal{P}(C_i^j)}, \quad \forall (x, y) \in I_i\hspace{0.5ex},
\label{eq:comap_i}
\end{equation}
where $\delta_{(x, y) \in \mathcal{P}(C_i^j)}$ is an indicator variable returning $1$ if pixel $(x, y)$ in $I_i$ has a correspondence match in $I_j$, and $0$ otherwise. Here, $\mathcal{P}(C_i^j)$ represents the set of pixel coordinates in $I_i$ that have correspondences in $I_j$.

The resulting covisibility map $M_i$ captures pixel-wise covisibility counts ranging from $0$ to $n - 1$, indicating the frequency with which a pixel appears across views, as shown in Fig.~\ref{fig:teaser}, with values rescaled from $0$ (black) to $255$ (white) for visualization purposes. Unlike DyCheck~\citep{gao2022dynamic}’s covisibility mask, which only indicates pixel visibility, our covisibility map provides precise covisibility details (\ie, covisible view count) at the pixel level. This map supports both the initial point cloud update (Sec.~\ref{subsec:InitPCL_update}) and the weighted supervision method (Sec.~\ref{subsec:comap_based_supervision}).

\subsection{Enhancing Initial Point Clouds}
\label{subsec:InitPCL_update}

In this subsection, we explain how to improve initial point clouds for both low-uncertainty multi-view and high-uncertainty mono-view regions. To enhance initial point clouds in \textbf{low-uncertainty multi-view regions} where sparse SfM points are available, we integrate a triangulated set of 3D points, $P_T$, derived from dense correspondences obtained from MASt3R~\citep{mast3r_arxiv24}, to complement COLMAP’s point cloud, $P_C$. This approach enhances geometrically plausible structures, especially in regions where COLMAP’s keypoint-based matching is sparse. For each view $I_i$, we generate $P_T$ by triangulating dense correspondences $C_i^{\{j\}}$ from view $I_i$ to other views $I_{\{j\}}$ (\eg, two or three correspondences in a 3-view case). To avoid redundancy, correspondences already reconstructed from other views are excluded from subsequent triangulations.

\begin{figure}[t!]
\centering
\includegraphics[width=0.995\linewidth]{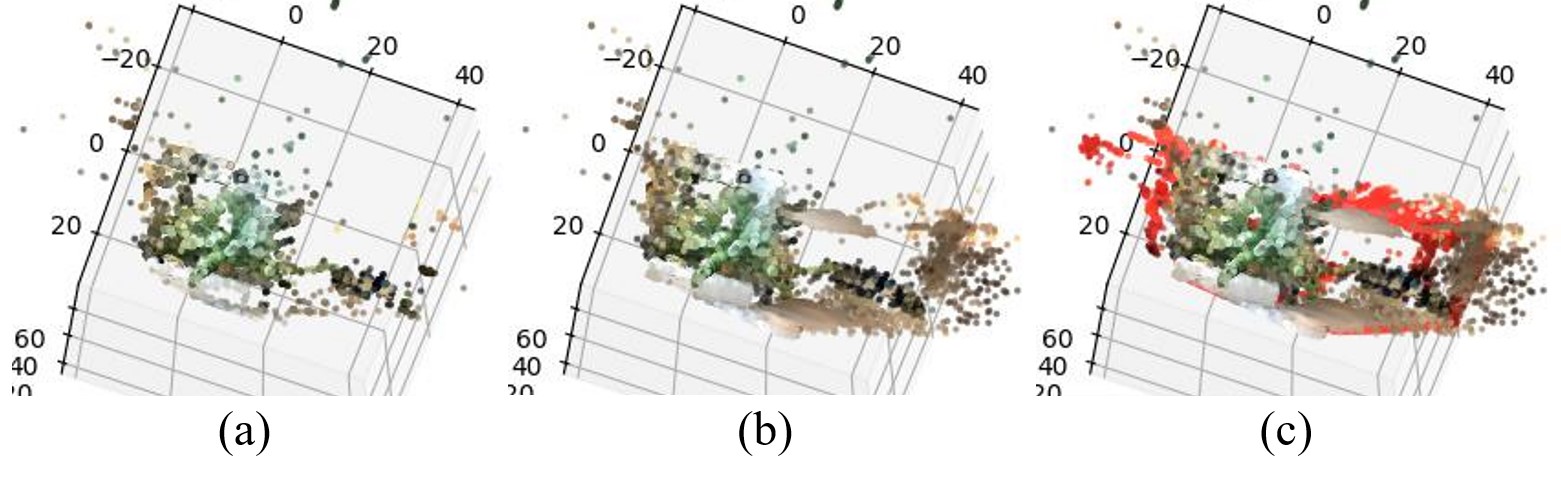}
\caption{Examples of enhanced initial PCLs from the Fern scene in the LLFF dataset. (a) Initial SfM points are first (b) updated through triangulation using dense correspondences and COLMAP camera poses, and then (c) fully refined to include mono-view regions (mono-view updates are colored in red for visualization).}
\label{fig:Init_PCL_update}
\end{figure}

Formally, we define the updated point cloud $P_\text{u}$ as:
\begin{equation}
    P_\text{u} = P_C \cup \{ p_T \in P_T \mid D(p_T, P_C) > \epsilon \}\hspace{0.5ex},
\label{eq:update_init_pcl_covisible_region}
\end{equation}
where $D(p_T, P_C)$ is the distance from a point $p_T$ in $P_T$ to its nearest neighbor in $P_C$, and $\epsilon$ is a threshold ensuring that only points sufficiently distant from existing points in $P_C$ are added. This selective integration allows the triangulated points to supplement COLMAP’s sparse initial point cloud in the same coordinate system, filling gaps in low-uncertainty regions and improving overall density in covisible areas, as shown in Fig.~\ref{fig:Init_PCL_update}.

In \textbf{high-uncertainty mono-view regions}, we use monocular depth estimation~\citep{HuTPAMI24Metric3Dv2} to generate a preliminary set of 3D points, $P_{d}$, by unprojecting depth estimates using the provided camera intrinsics. This set is defined as $P_{d} = P_{d}^{\text{low}} \cup P_{d}^{\text{high}}$, where $P_{d}^{\text{low}}$ comprises points unprojected from multiview regions (where $M_i \geq 1$), and $P_{d}^{\text{high}}$ comprises points unprojected from mono-view regions (where $M_i = 0$). To align these randomly scaled unprojected points with triangulated points from multiview geometry, we learn an anisotropic scaling transformation $f_{\text{scale}}$, using covisible points from $P_u^{\text{low}} \subseteq P_u$ as ground truth and $P_{d}^{\text{low}}$ as input. This transformation applies scaling parameters learned through linear regression to match the multiview-derived geometry: 
\begin{equation}
    f_{\text{scale}}(P_{d}^{\text{low}}) \approx P_u^{\text{low}}\hspace{0.5ex}.
\end{equation}
For consistency, we define $P_u = P_u^{\text{low}} \cup P_u^{\text{high}}$, where $P_u^{\text{low}}$ is selected based on covisibility: 
\begin{equation}
    P_u^{\text{low}} = \{ p \mid M_i(\pi(P_u, \mathbf{H}_{i})) \ge 1 \}\hspace{0.5ex},
\label{eq:extracting_multiview_points}
\end{equation}
where $\pi(P_u, \mathbf{H}_{i})$ denotes the projection of $P_u$ using the camera transformation matrix $\mathbf{H}_{i}$, yielding the image coordinates $(x, y)$ for a given camera viewpoint and ensuring inclusion of only points visible in multiple views.

After training, we apply $f_{\text{scale}}$ to the randomly scaled unprojected points from depth estimates in high-uncertainty mono-view regions:
\begin{equation}
    P_s^{\text{high}} = f_{\text{scale}}(P_{d}^{\text{high}})\hspace{0.5ex}.
\end{equation}
This transformation aligns the randomly scaled points $P_{d}^{\text{high}}$ to the scale of COLMAP’s triangulated points $P_s^{\text{high}}$, ensuring spatial consistency across regions. The final point cloud, $P_{\text{final}}$, is defined as:
\begin{equation}
    P_{\text{final}} = P_u^{\text{low}} \cup P_s^{\text{high}}\hspace{0.5ex},
\end{equation}
resulting in enhanced initial point cloud density and accuracy across both multiview and mono-view regions, as shown in Fig.~\ref{fig:Init_PCL_update}, thus supporting cohesive scene reconstruction in sparse view synthesis.

\subsection{Covisibility Map-Based Weighted Supervision}
\label{subsec:comap_based_supervision}

\begin{figure}[t!]
\centering
\includegraphics[width=0.995\linewidth]{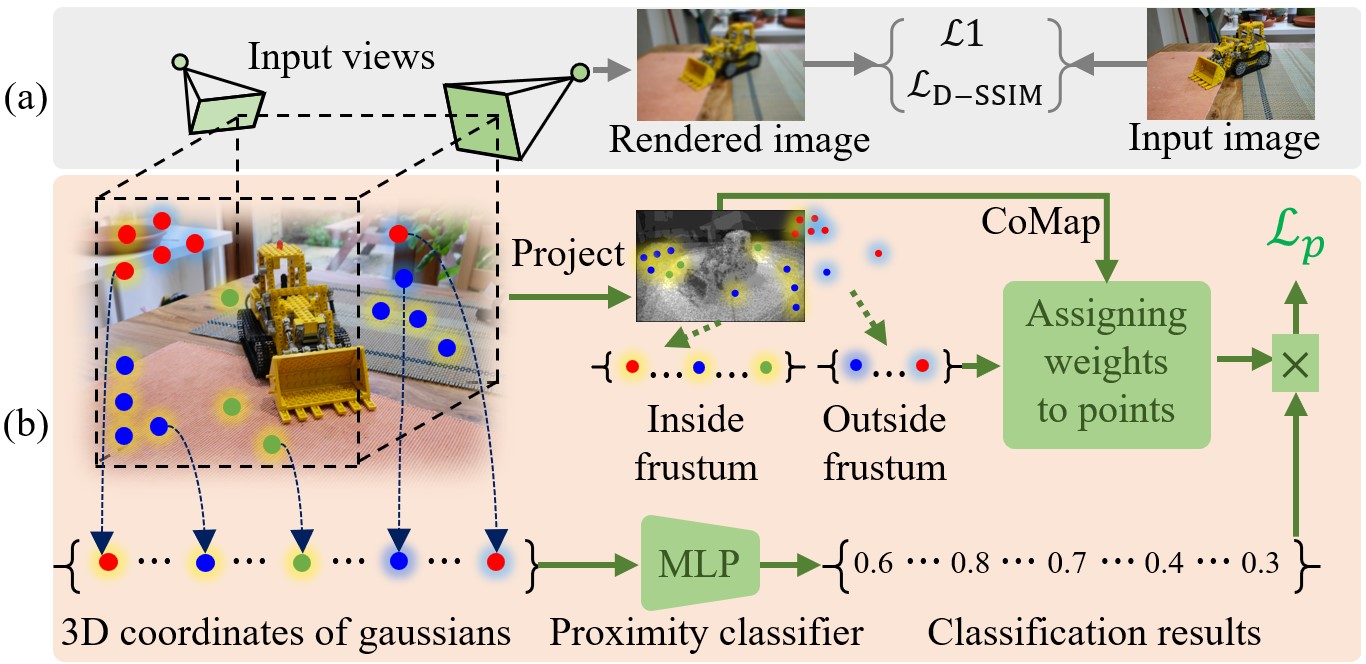}
\caption{Alongside (a) the conventional reconstruction loss $\mathcal{L}1$, we add (b) a weighted proximity loss $\mathcal{L}_{p}$ that applies classifier results with assigned weights based on Gaussian projections onto the covisibility map and their positions relative to the frustum.}
\label{fig:proximity_loss}
\end{figure}

We propose a covisibility map-based supervision strategy that leverages the enhanced initial point cloud, $P_{\text{final}}$, and the covisibility map $M_i$ to adaptively supervise regions based on uncertainty levels represented by covisibility counts, as shown in Fig.~\ref{fig:proximity_loss}. This approach improves conventional sparse view Gaussian splatting methods~\citep{Zhang_2024_ECCV_CorGS, Zhu_2024_ECCV_FSGS} by introducing region-specific weighting across covisibility levels, $M_i = \{0:n-1\}$, and by supervising Gaussians located outside the training view frustum with a scene-level weight based on the average covisibility score $S = \text{AVG}(M_{[1:n]})$.

Our approach integrates seamlessly with the latest 3DGS-based sparse view synthesis methods~\citep{Zhang_2024_ECCV_CorGS, Zhu_2024_ECCV_FSGS} by adding a proximity loss term, weighted using inverse proximity scores predicted by a three-layer MLP classifier trained on $P_{\text{final}}$ (see details in the supplementary materials).

\textbf{Proximity loss with adaptive weighting.} To implement this approach, we first train an MLP classifier $f_{p}$ on $P_{\text{final}}$, labeling these curated scene 3D points as positive (representing the original scene geometry) and randomly distributed points, $P_{\text{random}}$, constrained to remain distant from $P_{\text{final}}$, as negative. The classifier $f_{p} : \mathbb{R}^3 \to [0, 1]$ outputs a proximity score $s$ for each 3D point $p$:
\begin{equation}   
    s = f_{p}(p) \in [0, 1]\hspace{0.5ex},
\label{eq:proximity_classifier}
\end{equation}
where $s$ approaches 1 for points close to the scene geometry and 0 for distant points. Using batch processing, we compute the proximity scores efficiently for the entire Gaussian set, $G$, and define the proximity loss $\mathcal{L}_{p}$ as:
\begin{equation}   
    \mathcal{L}_{p}=\frac{1}{|G|}  \sum_{g \in G} \left( \chi(g) w_{\text{in}} + (1 - \chi(g)) w_{out} \right) \cdot \left(1 - s\right) \hspace{0.5ex},
\label{eq:proximity_loss_term}
\end{equation}
where $g$ denotes the 3D position of a Gaussian in the set $G$, representing the complete Gaussian set. The indicator function $\chi(g)$ is defined as:
\begin{equation}   
    \chi(g) =
    \begin{cases}
    1, & \text{if } \pi(g, \mathbf{H}_i) \in [0, \mathsf{W}) \times [0, \mathsf{H}) \\
    0, & \text{otherwise}
    \end{cases}
    \hspace{0.5ex},
\label{eq:indicator_fn}
\end{equation}
where $\pi(g, \mathbf{H}_i)$ is the projection of $g$ under camera transformation $\mathbf{H}_i$ onto the covisibility map $M_i$.

\textbf{Within the camera-view frustum.} We use the weight $w_{in}$ for Gaussians projected onto covisibility mask regions, defined as:
\begin{equation}   
        w_{\text{in}} = \frac{1}{M_i(\pi(g, \mathbf{H}_{i})) + 1}\hspace{0.5ex},
\label{eq:covisible_region_weights}
\end{equation}
where $M_i(\pi(g, \mathbf{H}{i}))$ is the covisibility count of $g$ in the covisibility map for the corresponding projection. This weighting ensures that high-covisibility regions, which are frequently supervised by other loss terms, receive weaker proximity-related supervision, while mono-view regions benefit from a more direct application of $\mathcal{L}_{p}$.

\textbf{Outside the camera-view frustum.} In conventional 3DGS-based methods, only Gaussians within the view frustum are supervised per iteration. However, for scenes with high covisibility scores ($S > 0.7$), we extend supervision to Gaussians located outside the view frustum, applying a scene-level weight $w_{out}$ for these points:
\begin{equation}
    w_{out} = \max\left( 0, \frac{S - 0.7}{0.3} \right)\hspace{0.5ex},
\label{eq:outside_region_weights}
\end{equation}
which scales linearly from 1 to 0 as $S$ ranges from 1 to 0.7. This approach improves supervision outside the view frustum in high-covisibility score scenes, ensuring comprehensive guidance while making efficient use of the proximity classifier. 

\subsection{Objective Function} 
\label{subsec:objective_function}

Our proposed CoMapGS enhances the objective function by introducing a proximity loss term, effectively complementing the function while preserving the original strategies of state-of-the-art methods, such as Co-pruning and Pseudo-view Co-regularization from CoR-GS~\citep{Zhang_2024_ECCV_CorGS}. The full objective function is expressed as:
\begin{equation}
    \mathcal{L} = (1 - \lambda) \mathcal{L}1(I, I^{*}) + \lambda \mathcal{L}_{\text{D-SSIM}}(I, I^{*}) + \mathcal{L}_{p}\hspace{0.5ex},
\end{equation}
where $I$ is the rendered image, supervised against ground truth $I^{*}$, $\mathcal{L}1$ is the L1 reconstruction loss, and $\mathcal{L}_{\text{D-SSIM}}$ represents the D-SSIM loss term. The balancing parameter $\lambda$ controls the weighting between L1 and D-SSIM, in line with standard Gaussian splatting objectives. The proximity loss term $\mathcal{L}_{p}$, defined in Eq.~\ref{eq:proximity_loss_term}, adds adaptive supervision to improve geometry alignment.

Our proposed approach extends seamlessly to other recent sparse view synthesis methods, such as FSGS~\citep{Zhu_2024_ECCV_FSGS}, by incorporating $\mathcal{L}_{p}$ into their objective function while preserving their original strategies, as reported in Table~\ref{tab:llff_comparison}.

\section{Experimental Results}
\label{sec:experiments}

\begin{table*}[!ht]
\centering
\caption{Quantitative results on LLFF with 3, 6, and 9 training views. The best, second-best, and third-best entries are marked in red, orange, and yellow, respectively. $^{\dagger}$ indicates results reproduced by us using the released code base.}
\label{tab:llff_comparison}
\begin{tabular}{ l | c c c | c c c | c c c }
\toprule
Method & \multicolumn{3}{c|}{PSNR $\uparrow$} & \multicolumn{3}{c|}{SSIM $\uparrow$} & \multicolumn{3}{c}{LPIPS $\downarrow$} \\
       & 3-view & 6-view & 9-view & 3-view & 6-view & 9-view &3-view & 6-view & 9-view  \\
\midrule
DietNeRF~\citep{Jain_2021_ICCV_DietNeRF} & 14.94 & 21.75 & 24.28 & 0.370 & 0.717 & 0.801 & 0.496 & 0.248 & 0.183 \\
Mip-NeRF~\citep{barron2021mipnerf} & 16.11 & 22.91 & 24.88 & 0.401 & 0.756 & 0.826 & 0.460 & 0.213 & 0.160 \\
RegNeRF~\citep{Niemeyer2021Regnerf} & 19.08 & 23.10 & 24.86 & 0.587 & 0.760 & 0.820 & 0.336 & 0.206 & 0.161 \\
SimpleNeRF~\citep{somraj2023simplenerf} & 19.24 & 23.05 & 23.98 & 0.623 & 0.737 & 0.762 & 0.375 & 0.296 & 0.286 \\
FreeNeRF~\citep{Yang2023FreeNeRF} & 19.63 & 23.73 & 25.13 & 0.612 & 0.779 & 0.827 & 0.308 & 0.195 & 0.160 \\
SparseNeRF~\citep{Wang_2023_ICCV_SparseNeRF} & 19.86 & 23.26 & 24.27 & 0.714 & 0.741 & 0.781 & 0.243 & 0.235 & 0.228 \\
DiffusioNeRF~\citep{Wynn_2023_CVPR_DiffusionNeRF} & 20.13 & 23.60 & 24.62 & 0.631 & 0.775 & 0.807 & 0.344 & 0.235 & 0.216 \\
ReconFusion~\citep{Wu_2024_CVPR} & \cellcolor{red!25} 21.34 & 24.25 & 25.21 & \cellcolor{yellow!25} 0.724 & 0.815 & 0.848 & 0.203 & 0.152 & 0.134 \\
\midrule
DNGaussian~\citep{Li_2024_CVPR_DNGaussian} & 19.12 & 22.18 & 23.17 & 0.591 & 0.755 & 0.788 & 0.294 & 0.198 & 0.180 \\ 
3DGS~\citep{kerbl3Dgaussians} & 19.22 & 23.80 & 25.44 & 0.649 & 0.814 & 0.860 & 0.229 & 0.125 & 0.096 \\
FSGS~\citep{Zhu_2024_ECCV_FSGS}$^{\dagger}$ & 20.458 & 24.694 & 26.006 & 0.713 & 0.839 & 0.875 & 0.205 & 0.121 & 0.095 \\
CoR-GS~\citep{Zhang_2024_ECCV_CorGS}$^{\dagger}$ & 20.473 & \cellcolor{yellow!25} 24.777 & \cellcolor{yellow!25} 26.475 & 0.717 & \cellcolor{orange!25} 0.844 & \cellcolor{orange!25} 0.881 & \cellcolor{yellow!25} 0.199 & \cellcolor{orange!25} 0.116 & \cellcolor{orange!25} 0.086 \\
\midrule
~\citep{Zhu_2024_ECCV_FSGS}$^{\dagger}$ + CoMapGS (ours) & \cellcolor{yellow!25} 20.891 & \cellcolor{orange!25} 24.808 & \cellcolor{orange!25} 26.292 & \cellcolor{orange!25} 0.737 & \cellcolor{orange!25} 0.844 & \cellcolor{yellow!25} 0.879 & \cellcolor{orange!25} 0.190 & \cellcolor{yellow!25} 0.117 & \cellcolor{yellow!25} 0.092 \\
~\citep{Zhang_2024_ECCV_CorGS}$^{\dagger}$ + CoMapGS (ours) & \cellcolor{orange!25} 21.105 & \cellcolor{red!25} 25.204 & \cellcolor{red!25} 26.731 & \cellcolor{red!25} 0.747 & \cellcolor{red!25} 0.854 & \cellcolor{red!25} 0.886 & \cellcolor{red!25} 0.182 & \cellcolor{red!25} 0.108 & \cellcolor{red!25} 0.082 \\
\bottomrule
\end{tabular}
\end{table*}

\begin{figure*}[!ht]
\centering
\includegraphics[width=0.995\linewidth]{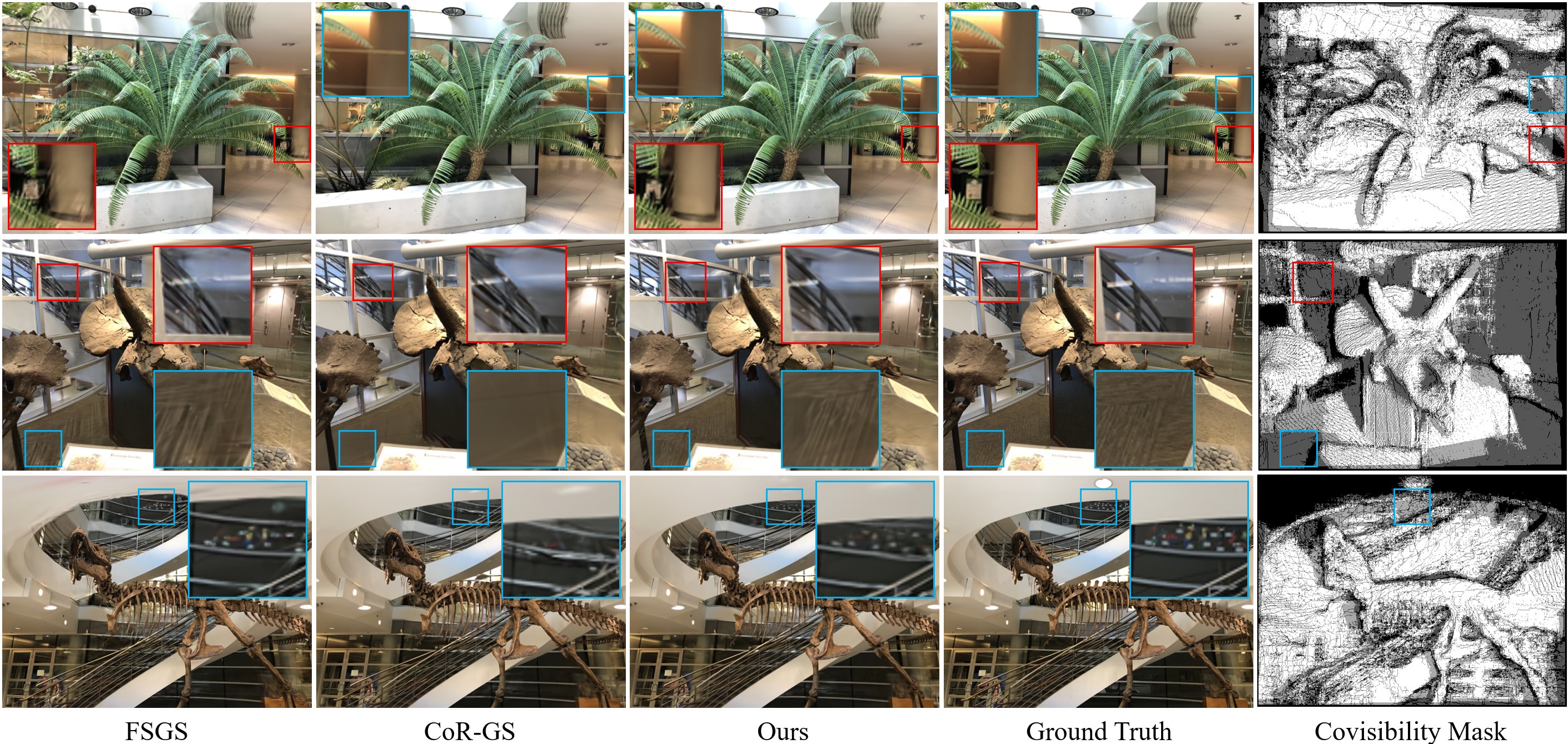}
\caption{Qualitative comparison on 3 training views from the LLFF dataset. We compare our results with FSGS~\citep{Zhu_2024_ECCV_FSGS} and CoR-GS~\citep{Zhang_2024_ECCV_CorGS}, demonstrating that our method achieves superior texture synthesis in novel views, particularly in high-uncertainty mono-view regions that appear only once in the training set.
}
\label{fig:comparison_on_llff}
\end{figure*}

\begin{table*}[ht]
\centering
\caption{Quantitative results on Mip-NeRF360 with 12 and 24 training views. The best, second-best, and third-best entries are marked in red, orange, and yellow, respectively. $^{\dagger}$ indicates results reproduced by us using the released code base.}
\label{tab:360_comparison}
\begin{tabular}{ l | c c | c c | c r }
\toprule
Method & \multicolumn{2}{c|}{PSNR $\uparrow$} & \multicolumn{2}{c|}{SSIM $\uparrow$} & \multicolumn{2}{c}{LPIPS $\downarrow$} \\
       & 12-view & 24-view & 12-view & 24-view & 12-view & 24-view  \\
\midrule
3DGS~\citep{kerbl3Dgaussians} & 18.52 & 22.80 & 0.523 & 0.708 & \cellcolor{yellow!25}0.415 & 0.276  \\
FSGS~\citep{Zhu_2024_ECCV_FSGS} & \cellcolor{yellow!25}18.80 & \cellcolor{yellow!25}23.28 & \cellcolor{yellow!25}0.531 & \cellcolor{yellow!25}0.715 & 0.418 & \cellcolor{yellow!25}0.274  \\
CoR-GS~\citep{Zhang_2024_ECCV_CorGS}$^{\dagger}$ & \cellcolor{orange!25}19.162 & \cellcolor{orange!25}23.321 & \cellcolor{orange!25}0.574 & \cellcolor{orange!25}0.729 & \cellcolor{orange!25}0.414 & \cellcolor{orange!25}0.271  \\
~\citep{Zhang_2024_ECCV_CorGS}$^{\dagger}$ + CoMapGS (ours) & \cellcolor{red!25}19.680 & \cellcolor{red!25}23.462 & \cellcolor{red!25}0.591 & \cellcolor{red!25}0.734 & \cellcolor{red!25}0.394 & \cellcolor{red!25}0.264  \\
\bottomrule
\end{tabular}
\end{table*}

\begin{figure*}[!ht]
\centering
\includegraphics[width=0.99\linewidth]{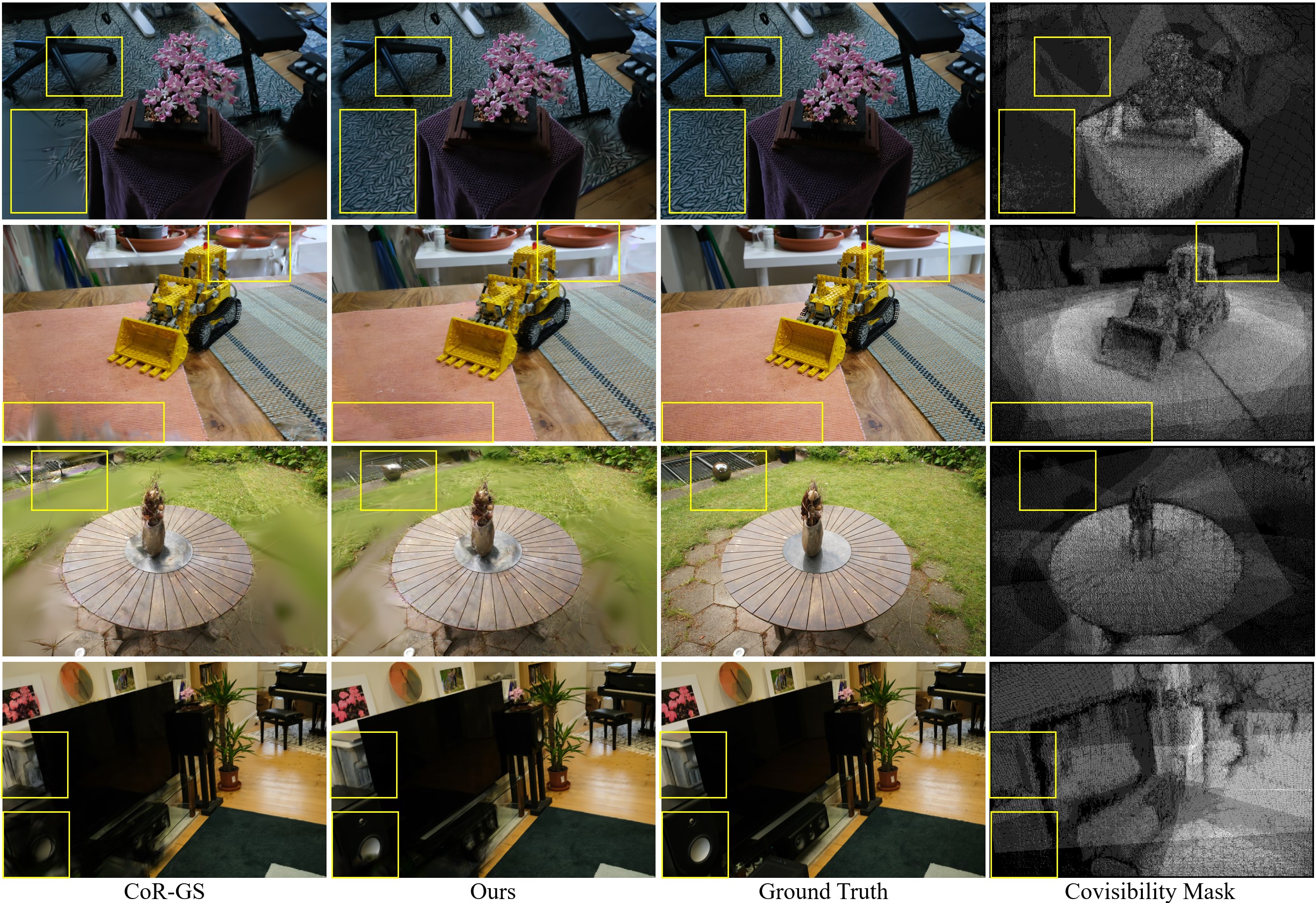}
\caption{Qualitative comparison on 12 training views from the Mip-NeRF 360 dataset. Mip-NeRF 360 highlights diverse covisibility levels, with high-uncertainty regions sparsely captured. CoMapGS effectively synthesizes these regions, while CoR-GS struggles to recover textures.}
\label{fig:comparison_on_360}
\end{figure*}

\subsection{Setup}
\textbf{Datasets.} We conducted experiments on two benchmark datasets: LLFF~\citep{mildenhall2019llff}, using 3, 6, and 9 training views with resolutions downsampled to 8×, and Mip-NeRF 360~\citep{Barron_2022_CVPR}, using 12 and 24 training views per scene with resolutions downsampled to 4×. We adhered to the standard training and testing conventions~\citep{Zhang_2024_ECCV_CorGS} for datasets commonly used in sparse view synthesis.

\noindent\textbf{Comparison Baselines.} We compare CoMapGS with existing few-shot novel view synthesis methods, including FSGS~\citep{Zhu_2024_ECCV_FSGS} and CoR-GS~\citep{Zhang_2024_ECCV_CorGS}. As our method primarily enhances the performance of 3DGS-based approaches, we reproduced results for these methods. For other methods, we report the best published scores from prior works, as shown in Tables~\ref{tab:llff_comparison} and \ref{tab:360_comparison}.

\noindent\textbf{Evaluation Metrics.} We evaluate our results using PSNR, SSIM, and LPIPS metrics. 

\noindent\textbf{Implementation.} Training was conducted for 10k iterations on the LLFF dataset and 30k iterations on the Mip-NeRF 360 dataset, following the setup of 3DGS~\citep{kerbl3Dgaussians}. Consistent with experimental setups in FSGS~\citep{Zhu_2024_ECCV_FSGS} and CoR-GS~\citep{Zhang_2024_ECCV_CorGS}, CoMapGS was initialized with point clouds and precomputed camera poses from COLMAP.

\subsection{Comparisons}
\label{subsec:comparison1}

\textbf{LLFF.} Quantitative and qualitative results on the LLFF dataset are presented in Table~\ref{tab:llff_comparison} and Fig.~\ref{fig:comparison_on_llff}, respectively. We reproduced evaluation scores for the two baseline methods, FSGS~\citep{Zhu_2024_ECCV_FSGS} and CoR-GS~\citep{Zhang_2024_ECCV_CorGS}, to highlight the improvements over our reproduced results. The results show consistent improvements, ranking first across various training view counts, with PSNR placing second only to ReconFusion~\citep{Wu_2024_CVPR} in the 3-view case. This difference is likely due to diffusion-based sparse view synthesis methods, which are optimized for generating visually seamless colors that blend with surrounding regions. However, this approach may not accurately capture original patterns in mono-view or unseen regions—common in sparse view synthesis datasets, especially with fewer training views. Consequently, while we ranked second in PSNR, our method achieved substantially higher SSIM and lower LPIPS scores under comparable view settings, indicating improved structural fidelity and perceptual quality in the synthesized views.

\textbf{Mip-NeRF 360.} To validate CoMapGS’s applicability to broader scenes that capture large areas with outward-facing cameras, we evaluated our method using the Mip-NeRF 360 dataset. This dataset, which centers objects of interest with outward-facing cameras, captures the motivation of our approach by featuring distinct low-uncertainty, multiview regions alongside high-uncertainty, less-covisible regions within an image, as shown in Fig.~\ref{fig:teaser}. In these scenarios, COLMAP struggles to establish stable correspondences, leading to sparse initial point clouds that fail to adequately recover details in high-uncertainty regions, as shown in Fig.~\ref{fig:comparison_on_360}.

In contrast, CoMapGS begins with enhanced initial point clouds, preserving the scene’s original structure while benefiting from proximity loss supervision, resulting in improved texture representation over conventional methods, as shown in Fig.~\ref{fig:comparison_on_360}. Quantitative results in Table~\ref{tab:360_comparison} confirm significant improvements across metrics, particularly LPIPS, compared to recent state-of-the-art methods, which perform worse than their baselines (\eg, 3DGS~\citep{kerbl3Dgaussians}, FSGS~\citep{Zhu_2024_ECCV_FSGS}). In the 24-view case, while PSNR improvements may appear modest due to unseen regions in test views, the LPIPS gain highlights our method’s effectiveness in recovering textures from both low- and high-uncertainty regions in novel view.

\subsection{Ablation Study}
\label{subsec:ablation}

The ablation study in Table~\ref{tab:ablation_study} highlights that dense reconstruction in low-uncertainty (covisible) regions already provides significant improvement over baseline methods, underscoring the benefit of increased point density in covisible areas. Furthermore, adding points from mono-view regions to the enhanced point cloud yields additional gains, demonstrating the value of incorporating points from underrepresented, high-uncertainty regions in sparse view synthesis.

We also evaluated the impact of our proximity loss-based weighted supervision both with and without the initial point cloud enhancement strategy. Interestingly, starting from sparse COLMAP point clouds without updates showed limited improvement, likely because 3DGS-based methods tend to densify Gaussians near their original positions. However, when initialized with our enhanced point clouds, the benefits of weighted supervision became evident, delivering significant performance gains in both partial (multiview region) and full (multi- and mono-view region) update scenarios. This demonstrates that enhancing initial point clouds and incorporating proximity loss-based weighted supervision work synergistically to optimize 3DGS-based sparse view synthesis.

\begin{table}[!ht]
\centering
\caption{Ablation study of the proposed CoMapGS with six training views. $\bigtriangleup$ indicates enhancing initial point clouds only in the low-uncertainty region (where covisible count $\geq 2$), representing a straightforward application of dense point clouds generated by conventional methods such as MASt3R~\citep{mast3r_arxiv24}.}
\label{tab:ablation_study}
\begin{tabular}{ c c c c c }
\toprule
\multicolumn{2}{c}{Proposed method} & \multicolumn{3}{c}{LLFF} \\
Sec.~\ref{subsec:InitPCL_update} & Sec.~\ref{subsec:comap_based_supervision} & PSNR $\uparrow$ & SSIM $\uparrow$ & LPIPS $\downarrow$ \\
\midrule
\multicolumn{2}{c}{CoR-GS~\citep{Zhang_2024_ECCV_CorGS}$^{\dagger}$}  & 24.777 & 0.844  & 0.116 \\
         & \cmark             & 24.787 & 0.845 & 0.116 \\
\hdashline
$\bigtriangleup$ &                    & 24.90 & 0.849  & 0.112  \\
$\bigtriangleup$ & \cmark             & 25.153 & 0.854 & 0.109 \\
\hdashline
\cmark &                    & 25.076 & 0.852  & 0.109  \\
\cmark & \cmark             & \textbf{25.204} & \textbf{0.854}  & \textbf{0.108} \\
\bottomrule
\end{tabular}
\end{table}

\subsection{Discussion}
\label{subsec:discussions}

\textbf{Definition of sparse view synthesis.} Although sparse novel view synthesis is often defined by a specific number of training views (e.g., LLFF with 3, 6, or 9 views; Mip-NeRF with 12 or 24 views), we argue that the true challenge lies in reconstructing underrepresented regions with low covisibility counts rather than the total view count. This problem is inherently region-wise and imbalanced, where certain areas appear in only a few views, leading to sparse representation even when many views are available. For example, as DyCheck~\citep{gao2022dynamic} categorizes scene types based on effective multiview factors (EMFs), scenarios with high frame counts but low region-specific covisibility can still be considered sparse view cases. We address this by applying region-wise adaptive supervision based on covisibility counts, improving results, especially for imbalanced datasets like Mip-NeRF 360 (12 views). In such cases, central objects are densely captured, while outer areas show low covisibility due to the 360-degree, outward-facing capture setup. This issue is common in real-world scene capture, where uneven covisibility across views motivates balanced, covisibility-aware supervision.

\section{Conclusions}
\label{sec:conclusion}

This paper introduces CoMapGS, a Covisibility Map-based Gaussian Splatting approach designed to enhance sparse view synthesis methods by addressing region-specific uncertainty levels. We generate covisibility maps for a given non-continuous set of distant images, using them to identify low- and high-uncertainty regions and to improve point cloud initialization beyond conventional methods like COLMAP. By introducing covisibility-adaptive weighted supervision, CoMapGS effectively manages varying sparsity across both multiview and mono-view regions, consistently outperforming state-of-the-art 3DGS-based sparse novel view synthesis methods. Experimental results validate CoMapGS’s effectiveness across diverse, realistic scenes, including both indoor and outdoor environments and various scales of captured areas. CoMapGS demonstrates superiority in handling complex region-wise covisibility, highlighting its robustness and adaptability in challenging sparse view synthesis tasks.
{
    \small
    \bibliographystyle{ieeenat_fullname}
    \bibliography{main}
}

\clearpage
\setcounter{page}{1}
\maketitlesupplementary

\appendix

\section*{Supplementary Material Overview}

This supplementary document is structured as follows:
\begin{itemize}
  \item Section A: Validating Geometrical Correctness
  \item Section B: Supplementary Video Overview
  \item Section C: Addressing Concerns of the Proximity Loss
  \item Section D: Implementation and Runtime Efficiency
  \item Section E: Limitations
\end{itemize}

\section{Validating Geometrical Correctness}

\begin{figure*}[!ht]
\centering
\includegraphics[width=0.95\linewidth]{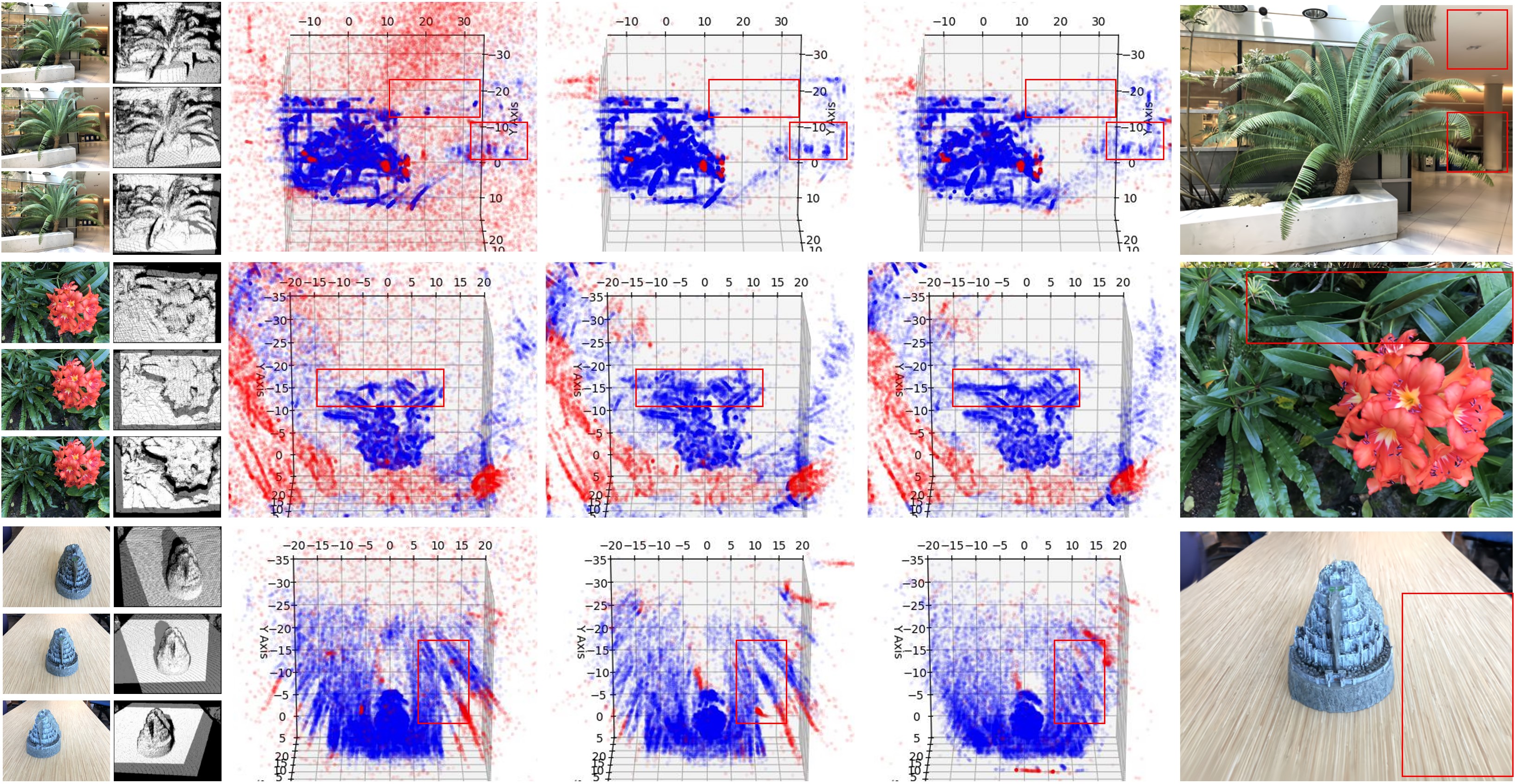} \\
\includegraphics[width=0.95\linewidth]{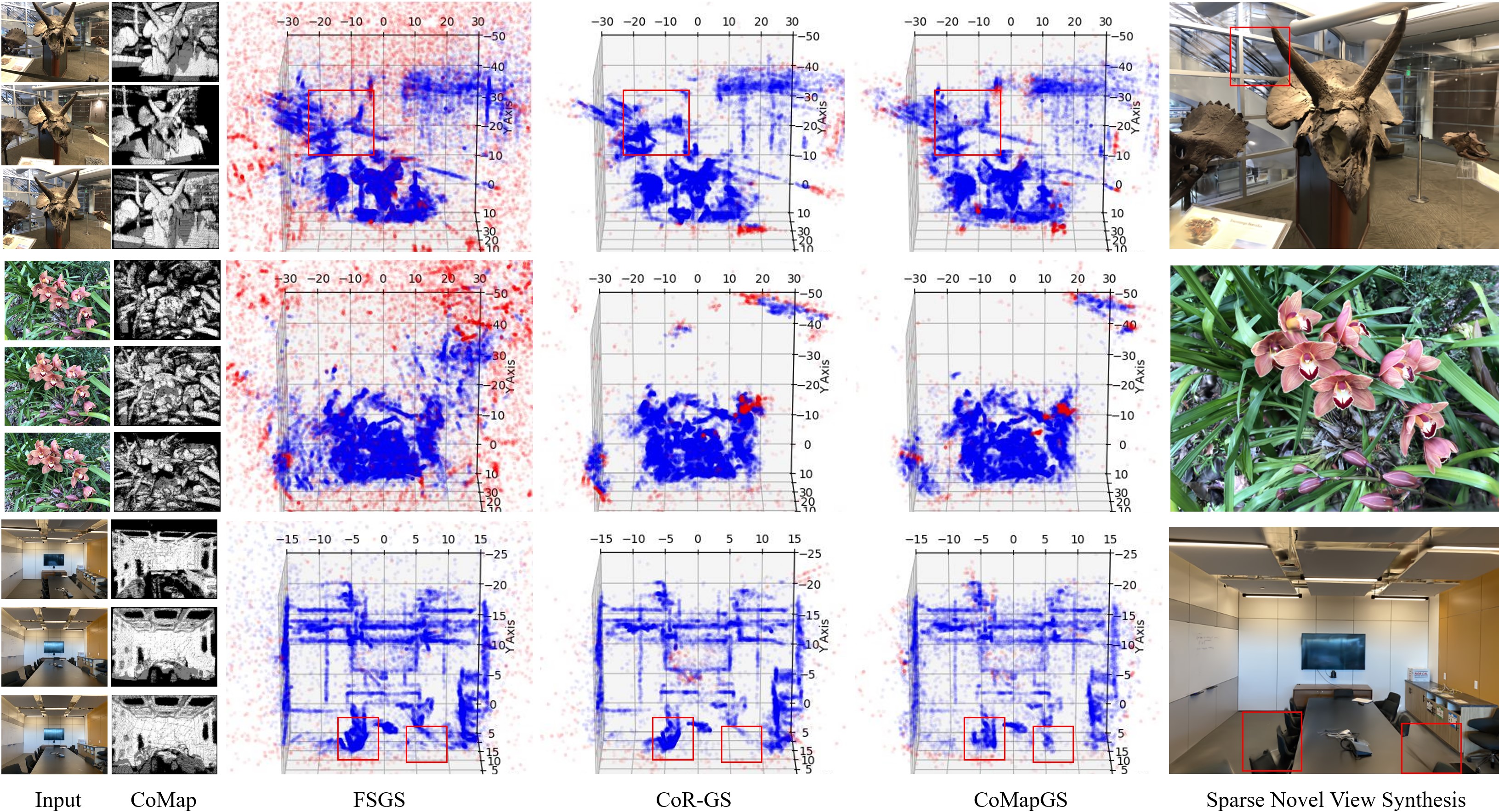}
\caption{Visualization of Gaussian positions after training on the LLFF dataset. FSGS shows overfitting with uncontrolled Gaussian proliferation, while CoR-GS strongly penalizes low-covisibility regions. CoMapGS achieves geometrically aligned Gaussians across both high- and low-covisibility regions. Blue and red points indicate proximity classification results (threshold 0.5), with proximity classifiers trained on 9 views for evaluation.}
\label{fig:sup_final_pcls_llff}
\end{figure*}

\begin{figure*}[!ht]
\centering
\includegraphics[width=0.95\linewidth]{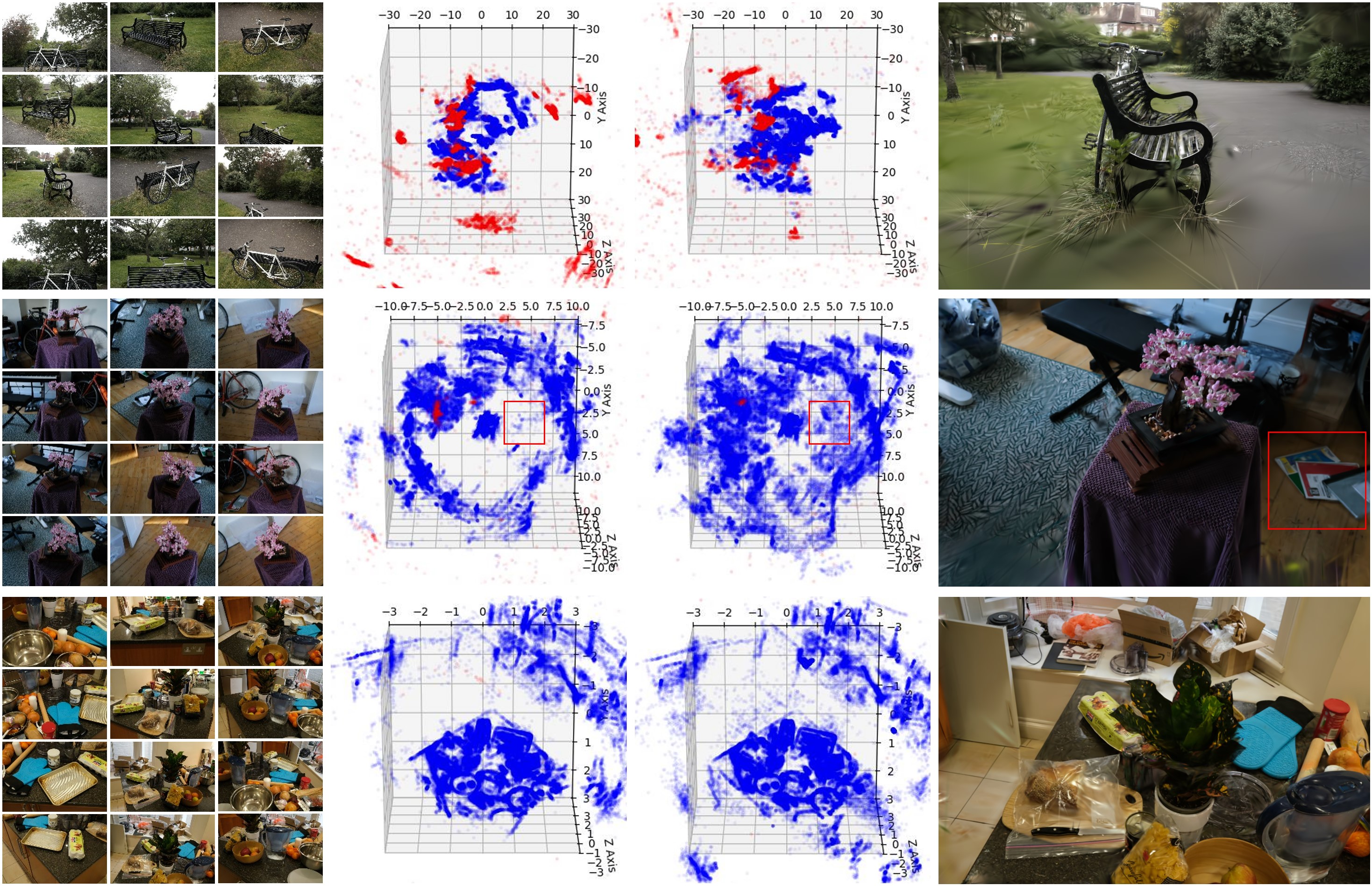} \\
\includegraphics[width=0.95\linewidth]{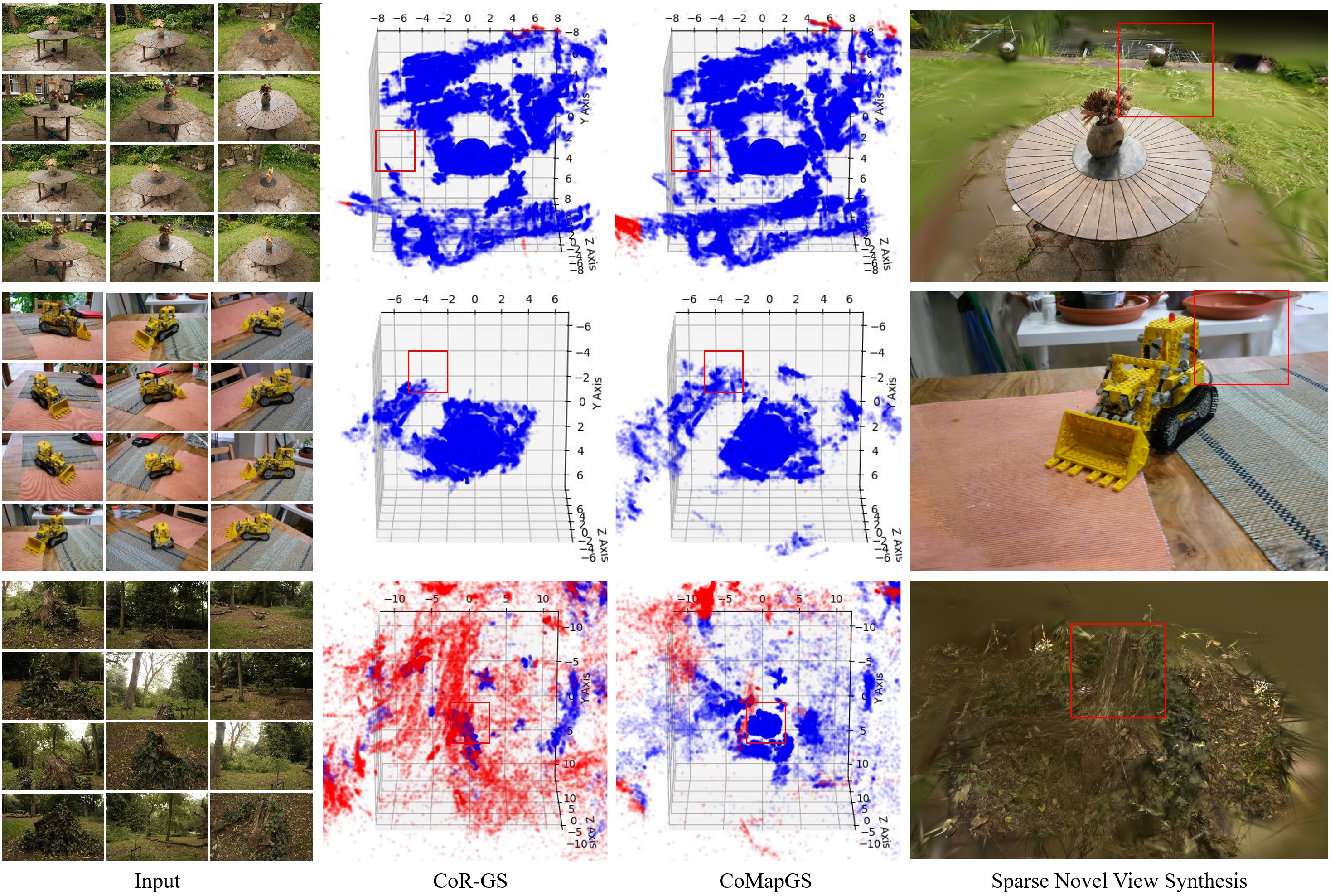}
\caption{Visualization of Gaussian positions after training on the Mip-NeRF 360 dataset. CoR-GS~\citep{Zhang_2024_ECCV_CorGS} penalizes low-covisibility regions strongly, leading to incomplete reconstructions. CoMapGS achieves balanced Gaussian distributions with improved geometric alignment across both high- and low-covisibility regions. Blue and red points indicate proximity classification results (threshold 0.5), with classifiers trained on 24 views for evaluation.}
\label{fig:sup_final_pcls_360}
\end{figure*}

To validate the geometrical correctness of the 3D Gaussian splatting results generated by CoMapGS, we analyze the reconstructed 3D point cloud positions after training. Fig.~\ref{fig:sup_final_pcls_llff} visualizes the completed set of Gaussian positions for the LLFF dataset trained with 3 views. To evaluate the proximity of the reconstructed geometry to the true scene, we applied a proximity classifier trained with 9 views, thus providing a broader reference set for assessing correctness.

\subsection{Comparative Analysis of the Latest Methods}

As shown in Fig.~\ref{fig:sup_final_pcls_llff}, the latest methods demonstrate different behaviors in sparse settings. Specifically:

\noindent \textbf{$\boldsymbol{\cdot}$ FSGS~\citep{Zhu_2024_ECCV_FSGS}} exhibits a tendency to overfit to pseudo-ground truth during the training process. This leads to an uncontrolled proliferation of Gaussians in high-uncertainty regions, often resulting in geometrically incorrect structures. The method’s reliance on dense Gaussian generation without effective geometric alignment contributes to artifacts, especially in mono-view regions.

\noindent \textbf{$\boldsymbol{\cdot}$ CoR-GS~\citep{Zhang_2024_ECCV_CorGS}} penalizes mono-view regions more strongly because the method is designed to retain Gaussians that are consistent across separate Gaussian models during training, ultimately prioritizing those that satisfy multiview geometry constraints. While this helps avoid overfitting in sparsely captured areas, it also leads to incomplete reconstructions or missing details in high-uncertainty regions. The lack of adaptive supervision across varying covisibility levels limits its ability to recover underrepresented structures.

\noindent \textbf{$\boldsymbol{\cdot}$} The proposed \textbf{CoMapGS} demonstrates significantly improved geometric alignment, as seen in the more coherent distribution of Gaussians across both high- and low-covisibility regions. By leveraging covisibility maps, CoMapGS applies adaptive supervision to ensure balanced updates for both multiview and mono-view regions. This is further supported by the proximity loss, which aligns Gaussians to the true scene structure without over-penalizing sparse areas.

\subsection{Insights from Proximity Classification}

In Fig.~\ref{fig:sup_final_pcls_llff}, the blue and red points denote the proximity classification results, with a threshold of 0.5. Blue points represent regions classified as closer to the true geometry, while red points indicate areas farther from the original structure. Key observations include: 

\noindent \textbf{$\boldsymbol{\cdot}$ Consistency in Multiview Regions}: In areas with high covisibility, CoMapGS effectively aligns Gaussians close to the ground truth, as evidenced by the predominance of blue points. This indicates that the adaptive weighting mechanism in multiview regions successfully guides reconstruction.

\noindent \textbf{$\boldsymbol{\cdot}$ Improvement in Mono-View Regions}: Unlike FSGS~\citep{Zhu_2024_ECCV_FSGS} and CoR-GS~\citep{Zhang_2024_ECCV_CorGS}, which either overfit or neglect mono-view regions, CoMapGS ensures that Gaussians in these high-uncertainty areas are plausibly positioned. The inclusion of proximity-based supervision mitigates structural divergence, which is critical for realistic novel view synthesis.

\noindent \textbf{$\boldsymbol{\cdot}$ Balanced Distribution}: CoMapGS prevents the over-concentration of Gaussians in highly covisible regions, addressing the imbalance commonly observed in baseline methods. This balanced distribution directly translates to improved reconstruction quality, particularly in underrepresented areas.

\subsection{Broader Implications}

The ability of CoMapGS to produce geometrically correct Gaussian distributions highlights its robustness and adaptability across varying levels of sparsity. The integration of covisibility-aware supervision ensures that even regions with minimal training views are adequately represented, making the method suitable for complex real-world scenes as shown in Fig.~\ref{fig:sup_final_pcls_360}.

\subsection{Conclusion}

The supplementary analysis presented here reinforces the findings in the main manuscript, demonstrating the effectiveness of CoMapGS in addressing the unique challenges of sparse view synthesis. By leveraging covisibility maps and proximity loss, the method achieves geometrically correct reconstructions, providing significant improvements over state-of-the-art 3DGS-based approaches.

\section{Supplementary Video Overview}

\noindent In addition to the visualizations and analysis presented in this document, a supplementary demo video is provided. The video includes:
\begin{enumerate}
    \item Visualizations of covisibility maps generated for both LLFF and Mip-NeRF 360 datasets.
    \item Enhanced initial point cloud updates guided by covisibility maps.
    \item Explanation of the covisibility map-based supervision approach.
    \item Qualitative comparisons with baseline methods, demonstrating the effectiveness of CoMapGS in recovering underrepresented regions.
\end{enumerate}

\clearpage
\section{Addressing Concerns of the Proximity Loss}

\subsection{Visual Results from Ablation Studies}

\begin{figure}[h]
  \centering
  \includegraphics[width=0.995\linewidth]{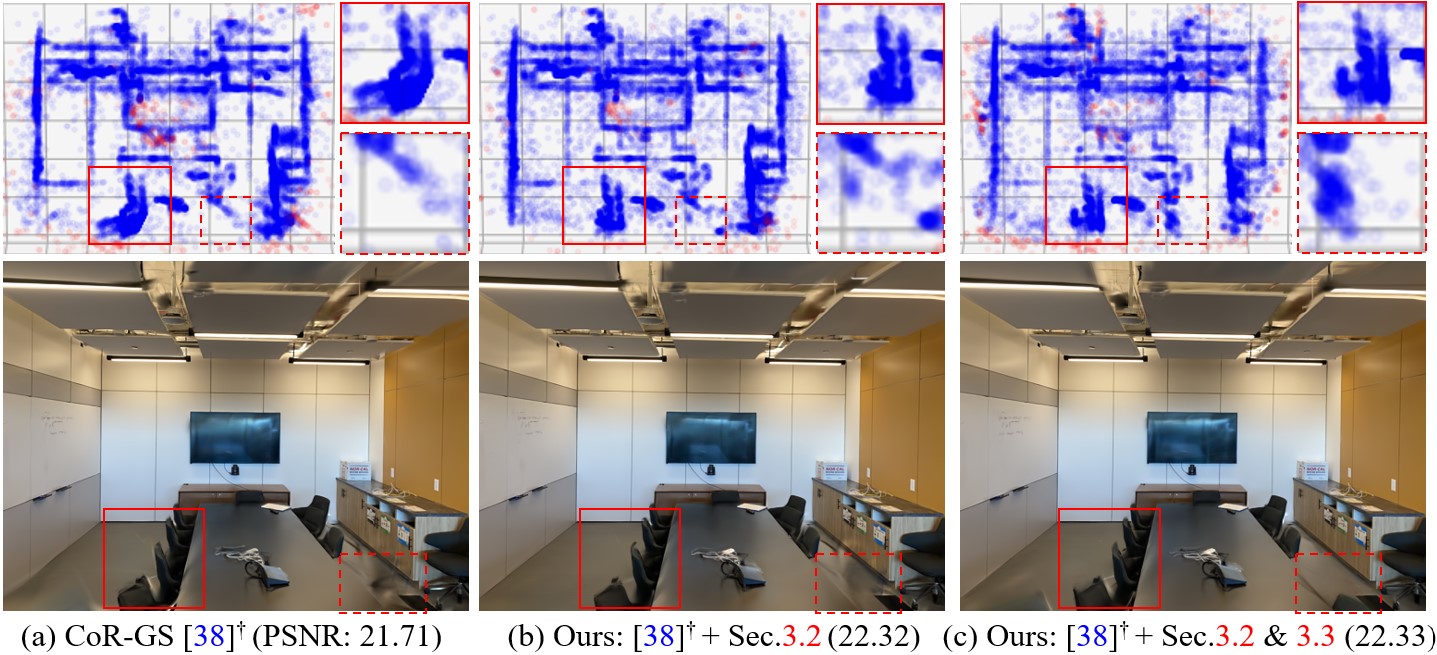}
   \caption{Ablation results with 3 training views, highlighting the effects of enhanced initial PCLs and proximity loss on novel view synthesis.
   }
   \label{fig:ablation_figure}
\end{figure}

Compared to Fig.~\ref{fig:ablation_figure}(a), which uses sparse point clouds (PCLs), Fig.~\ref{fig:ablation_figure}(b) shows that our enhanced PCLs correct the distorted geometry in the left box. Furthermore, Fig.~\ref{fig:ablation_figure}(c), which incorporates proximity loss, extends supervision beyond the frustum—preserving the overall 3D structure and preventing errors caused by incomplete supervision, as illustrated in the right dotted box (missing chair). As discussed in Sec.~\ref{subsec:ablation}, this highlights the complementary nature of proximity and photometric losses, leading to robust optimization across the entire scene.

\subsection{Expected Behavior Outside the Frustum}

The above ablation results also address concerns regarding the behavior of gaussians outside the frustum during training. Conventional 3DGS-based methods relying on photometric or depth losses adjust Gaussian parameters (3D coordinates, color, opacity, rotation, and scale) to improve rasterized color or depth images but do not explicitly guide 3D positions to converge. Proximity loss, computed independently of visibility via an MLP, penalizes wrongly located Gaussians while preserving geometrically correct points generated during initialization and densification, resulting in significantly improved geometric accuracy in the final reconstruction. 

\section{Implementation and Runtime Efficiency}

\subsection{Details of the Proximity Classifier}

The proximity classifier $f_{p}$ is a three-layer MLP represented as \((\text{Input: } \mathbb{R}^3) \xrightarrow{\text{FC+ReLU}} \mathbb{R}^{128} \xrightarrow{\text{FC+ReLU}} \mathbb{R}^{128} \xrightarrow{\text{FC+Sigmoid}} \mathbb{R}^{1}\), where each activation function follows its corresponding fully connected (FC) layer. The classifier is trained by minimizing binary cross-entropy loss using the Adam optimizer with a learning rate of 0.001 for 1,000 iterations per scene. Details on positive and negative data preparation are provided in Sec.~\ref{subsec:comap_based_supervision}.

\subsection{Time Complexity}

Table~\ref{tab:time_complexity} shows that the proposed CoMapGS introduces only a marginal overhead compared to CoR-GS~\citep{Zhang_2024_ECCV_CorGS}, primarily due to the preprocessing step $\alpha$ (+1.12 m) and the additional cost of running an MLP during training (+5.36 m). $\alpha$ includes CoMap generation (11.51 s), initial PCL updates (multiview: 23.20 s, mono-view: 0.24 s), and proximity classifier training (data generation: 8.41 s, training: 28.06 s).

\noindent Compared to RegNeRF~\citep{Niemeyer2021Regnerf}, a representative NeRF-based method which requires 69,768 training iterations, both CoR-GS~\citep{Zhang_2024_ECCV_CorGS} and our method require 10K iterations. While~\citep{Zhang_2024_ECCV_CorGS}+Full runs $f_p$ at every iteration, the additional training cost remains minimal (+5.36 m), keeping it computationally efficient. Inference runs in real-time. Overall, CoMapGS achieves higher efficiency than RegNeRF while staying comparable to~\citep{Zhang_2024_ECCV_CorGS}.

\begin{table}[h!]
\scriptsize
\centering
\caption{Time complexities for the Fern scene (3-view) in the LLFF.}
\label{tab:time_complexity}
\begin{tabular}{ c c c c c }
\toprule
Process & RegNeRF~\citep{Niemeyer2021Regnerf} &~\citep{Zhang_2024_ECCV_CorGS} &~\citep{Zhang_2024_ECCV_CorGS}+PCLs &~\citep{Zhang_2024_ECCV_CorGS}+Full \\
\midrule
Prep.(\textcolor{magenta}{C})$+\alpha$ & $\cdot$ & \textcolor{magenta}{C}OLMAP & \textcolor{magenta}{C}+35.35 s & \textcolor{magenta}{C}+72.22 s \\
Train & 307.12 m & 8.16 m & 7.51 m & 13.27 m \\
Render$/$Img & 9,226.33 ms & 3.16 ms & 2.56 ms & 3.30 ms \\
\bottomrule
\multicolumn{5}{l}{\scriptsize{* m = minutes, s = seconds, ms = milliseconds}} \\
\end{tabular}
\end{table}

\begin{figure}[!b]
  \centering
  \includegraphics[width=0.995\linewidth]{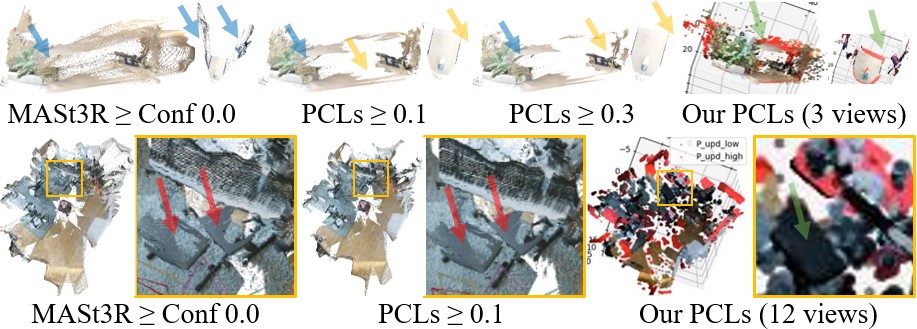}
   \caption{Illustration of how reprojection error validation robustly prevents \textcolor{green}{$\downarrow$} MASt3R errors (e.g., noisy PCLs \textcolor{blue}{$\downarrow$}, missing details \textcolor{orange}{$\downarrow$}, and misalignment \textcolor{red}{$\downarrow$}) when registering 3D points into COLMAP camera poses.}
   \label{fig:reliance_on_mast3r}
\end{figure}

\section{Limitations}

\subsection{MASt3R and Depth Estimation Dependence}

Our method relies on MASt3R and depth estimation, which can potentially propagate errors. To mitigate this, we register triangulated 3D points (using MASt3R’s dense correspondences) into COLMAP’s optimized camera poses with reprojection error validation ($\leq$ $2$ pixels), as shown in Fig.~\ref{fig:reliance_on_mast3r}. This prevents misaligned geometry across views and eliminates the need for manual confidence threshold tuning when using MASt3R’s predicted point clouds directly. While depth estimation errors may affect mono-view regions; however, their limited presence in the training set minimizes the impact. Additionally, we adjust the proximity loss weight (Eq.~\ref{eq:outside_region_weights}) based on the scene-level covisibility score $S$ to further reduce error propagation from depth predictions.

\subsection{Limitations on DTU Dataset}

The DTU dataset focuses on masked regions captured in an artificial, black, and structureless environment, where both depth prediction and COLMAP camera pose estimation often fail due to a lack of contextual information. While CoMapGS performs well in generic, real-world scenarios, it is less effective in controlled settings—or in scenes where the background is predominantly sky and thus geometrically projected to infinity. In such cases, objects are centrally positioned under controlled lighting, and the scene geometry is highly artificial, posing challenges for CoMapGS. Consequently, high-certainty mono-view regions may contain invalid geometry, increasing the likelihood of depth estimation errors. This, in turn, hinders the enhancement of initial point clouds and reduces the effectiveness of the proximity loss. 

Nevertheless, our tests on a subset of scenes from the DTU dataset indicate that performance does not degrade and remains comparable to the baseline CoR-GS. However, we were unable to run COLMAP successfully on the full DTU dataset and thus could not reproduce the results reported by recent methods. Preliminary experiments further suggest that synthetic or highly constrained environments may present similar limitations.

\subsection{Comparisons with NeRF-based Methods}

\begin{figure}[!h]
  \centering  \includegraphics[width=0.995\linewidth]{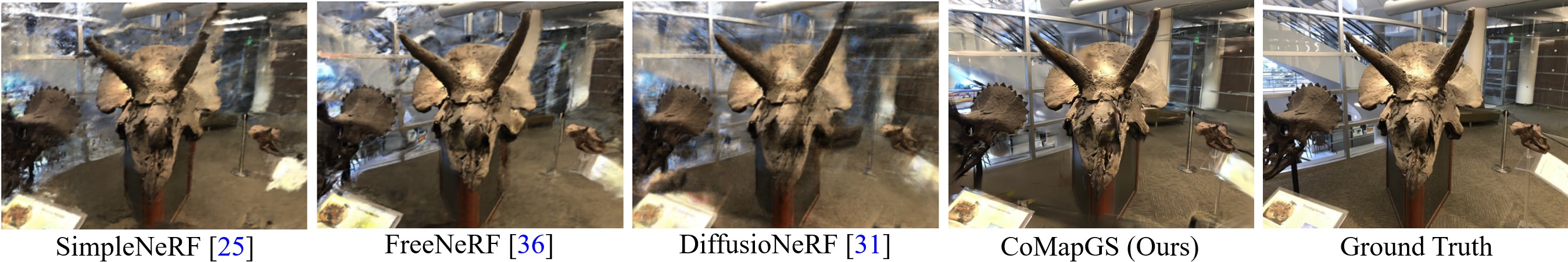}
   \caption{Visual comparisons with the latest NeRF-based methods.}
   \label{fig:NeRF_comp_figure}
\end{figure}

While our paper does not include an extensive set of qualitative comparisons with the latest NeRF-based methods, Fig.~\ref{fig:NeRF_comp_figure} presents a representative example. It highlights qualitative differences between our method and recent NeRF-based approaches, illustrating contrasts in rendering quality and geometric fidelity under sparse-view conditions.

\end{document}